# Semantic Alignment Between Normative Theories of Ethics and the European Union Artificial Intelligence Act: A Transformer-Based Semantic Textual Similarity Analysis

Mehmet Murat Albayrakoglu

Department of Management Information Systems
Isik University, Istanbul, Turkey
E-mail: murat.albayrakoglu@isikun.edu.tr
ORCID: 0000-0002-5057-5641

Mehmet Nafiz Aydin

Department of Management Information Systems
Bogazici University, Istanbul, Turkey
E-mail: mehmetn.aydin@bogazici.edu.tr
ORCID: 0000-0002-3995-6566

## Abstract

The European Union Artificial Intelligence (EU AI) Act, which explicitly references fundamental rights and ethical principles, is a comprehensive regulatory framework for governing Artificial Intelligence (AI) systems. This study examines the Act's possible moral grounding by analyzing the semantic alignment between three normative theories of ethics—virtue ethics, deontological ethics, and consequentialism—and the Act's regulatory language. Drawing on the philosophy of law, in which both antipositivist and some positivist accounts recognize an overlap between moral and legal norms, the relationship between the theories and the Act is treated as one of association. The encyclopedic descriptions of the theories serve as source documents, and the Act serves as the target document, analyzed in two target sections: its preamble and its statutory provisions. Semantic Textual Similarity (STS) is used as an indicator of association. The source documents are manually preprocessed to reduce semantic overlap among them. In all documents, each sentence is encoded by eight Transformer-based sentence encoders (SBERT, ALBERT, DistilBERT, RoBERTa, TinyBERT, MiniLM, E5-base, and GTE-base), and embeddings are aggregated to the document level by centroid cosine similarity, hierarchical sentence-block aggregation (HSBA; blocks of 3, 5, and 10 sentences), and mean pairwise sentence similarity. Because raw similarity scales differ across encoders, encoders are combined by voting on rankings. Deontological ethics shows the highest association with both target sections in all eight encoders and under every aggregation method, and the preamble is more closely associated with every source document than the provisions are. Consequentialism ranks second at the document level, but its position relative to virtue ethics depends on the granularity of aggregation.

## 1. Introduction

When legislation is devoid of ethical concern, deliberate or not, it may invite disaster because it lacks a moral basis to mitigate harm, again, intentional or not. Historically, many morally deficient laws have been enacted without due regard for ethical norms, and some have led to tragic consequences. (Radbruch (1946), 2006; Fuller, 1964). Among the examples are discriminatory legislation that favors a portion of society while condemning the rest, regulations that disregard potential environmental or health issues, laws that lead to the loss of historically and legally acquired property, and statutes that either disregard or outright violate fundamental human rights or are unable to moderate individual and societal conflicts.

Information Technology (IT) legislation is no exception to ethical concerns, as there are examples of adverse outcomes resulting from the avoidance or disregard of the moral aspects of laws and regulations. Surveillance laws can lead to violations of fundamental rights (e.g., Goitein, 2019; Meireles, 2022) or the heavy-handed treatment of individuals and groups by governments (e.g., Human Rights Watch, 2019). Some statutes have weaknesses due to segmentation, which can exacerbate the difficulties faced by disadvantaged groups (e.g., McCarthy, 2004), and to ambiguity, which can lead to indecisiveness or abuse (e.g., Kornbeck, 2021).

The European Union Artificial Intelligence (EU AI) Act has been subject to moral criticisms from the outset (e.g., Veale & Borgesius, 2021; Anderson, 2022). Despite being regarded as a significant step toward regulating Artificial Intelligence (AI) systems and emphasizing the fundamental rights (Musch et al., 2023), the risk-based approach taken to categorize and govern these systems is under scrutiny. One criticism of the Act is its focus on domain-specific criteria and technical compliance, without considering power asymmetries, fairness, and autonomy (Veale & Borgesius, 2021).

This multidisciplinary work aims to examine the relationship between ethics and the EU AI Act. Its interdisciplinary nature derives from the inclusion of ethics, philosophy of law, and linguistics. Among the contributing disciplines, computer science, and more specifically AI, was not only a subject of interest from ethical and legal perspectives, but it also provided the necessary tools and models to conduct analyses relevant to the study's aim.

The linguistic and computer science dimensions of the study were incorporated using Natural Language Processing (NLP), specifically Semantic Textual Similarity (STS), to estimate the extent of the association between ethics and law. To balance semantic expressiveness against resource constraints and environmental sensitivity, the study used eight Transformer-based sentence encoders ranging from about 12 to 110 million parameters, none larger than the standard base size of BERT.

The central proposition of the study is that STS indicates the association between the composition and meaning of a set of source documents and a target document, provided that a common theme or context links them. STS is not claimed to establish a causal or directional relationship, but to provide a comparative, text-related indicator of semantic alignment under an explicit contextual condition. Where the source documents predate the target document, that chronology is reported as context, but the association does not depend on it.

The documents whose content is compared with another document are called *source documents*, and the document with which they are compared is called the *target document*. These roles are analytical rather than causal: they fix the direction of comparison, not a direction of effect.

The source documents constitute the ethics dimension of the current work. For this study, three major, canonically distinct normative theories of ethics were selected as source documents. They are virtue ethics, deontological ethics, and consequentialism. These theories are normative because they prescribe principles for morally acceptable attitudes, decisions, and actions and provide criteria for moral judgment.

The target document is the EU AI Act of 2024, enacted to ensure that AI systems developed and used in the EU satisfy safety and transparency requirements without breaching ethical principles and fundamental rights. Within these confines, the Act allegedly aims to promote innovation and ensure the competitiveness of the EU institutions and businesses in AI. The Act is analyzed in two target sections, its preamble and its statutory provisions, to distinguish its intentional and operational groundings.

Prior interdisciplinary research in information ethics, AI governance, and technology policy has examined how normative ethical principles shape the design, governance, and societal implications of computational systems and regulatory frameworks (e.g., Floridi, 2002; Capurro, 2006; Stahl, 2012; Floridi et al., 2018; Jobin et al., 2019; Veale & Borgesius, 2021). These works highlight that ethical concepts are often embedded, explicitly or implicitly, within socio-technical systems, governance mechanisms, and legal discourse, particularly in the context of emerging AI regulation. Rather than engaging in prescriptive ethical judgment, such research frequently adopts analytical and interpretive approaches to examine how normative frameworks are reflected in institutional and policy texts.

Building on such an interdisciplinary perspective, the present study extends the discussion to regulatory language by computationally analyzing the semantic alignment between canonical normative theories of ethics and the European Union Artificial Intelligence Act, treating semantic similarity as an indicator of association rather than a direct measure of ethical or causal determination.

Textual documents are manifestations of their creators' thoughts and intentions. When documents address a common theme, repetition or systematic alignment of their compositional or semantic patterns indicates an association between them. In contrast, the absence of such patterns suggests no discernible association. Association, in this sense, makes no claim about the direction in which ideas traveled between the documents or their authors; it describes the extent to which the documents express overlapping meaning.

Because both the normative theories and the EU AI Act are textual artifacts reflecting the perspectives of philosophers and lawmakers, respectively, their semantic characteristics provide observable traces of the relationship between ethics and law. Given the historical depth and complexity of ethical traditions, constructing a detailed causal model of how ethical thought has shaped legislation is impractical. However, established philosophical discourse on the ethics-law relationship, in which antipositivist and some positivist accounts recognize that moral and legal norms overlap, supports the use of semantic textual similarity as an indicator of the association between theories of ethics and legal texts.

In the remainder of the paper, a literature review is presented that covers the key theoretical aspects of all contributing disciplines. Next, the contributions of each discipline are synthesized into a methodology in which eight Transformer-based sentence encoders compare each source document with each target section of the Act, namely, its preamble and statutory provisions. Sentence embeddings are aggregated to the document level with three methods: centroid cosine similarity, hierarchical sentence-block aggregation, and mean pairwise sentence similarity, and the encoders are combined by voting on the resulting rankings. Finally, the results are discussed, and conclusions are drawn about the work itself and its implications for the future.

## 2. Literature Review

The literature review examines prior work to contextualize the extent to which each normative theory relates to the EU AI Act. It covers the following: the relationship between ethics and law; normative theories, law, and the EU AI Act; the semantic association between texts; and Semantic Textual Similarity (STS) as an indicator of that association.

### 2.1. Moral Foundations of Law and Its Challenges to Natural Language Processing

Ethics of law, a branch of the philosophy of law, examines the relationship between ethics and law. Given the history of ethics and law, it is either impossible or prohibitively expensive to devise a detailed theoretical model that explicitly describes how ethical thought and legislation have been related over time. In the absence of such a model, a qualitative approach appears more appropriate; the discourse among philosophers of law on the relationship between ethics and law provides the necessary insight into the problem.

Despite the division between the philosophers of law about the relationship between ethics and law, current work relies mainly on the views of legal antipositivists: the law should be motivated by the moral concerns of its stakeholders, that is, voters, legislators, public administrators, and judges (Slote, 2001; Jowitt, 2022). The law serves as a bridge between its moral foundations and legal institutions and their practices (Postema, 2022).

Even some legal positivists agree that ethics and law overlap, as both are based on norms aimed at preventing harm and promoting good, the former among individuals and the latter within a society (Kramer, 2004). Such a thematic relationship between ethics and law also forms the basis for their shared context: dos and don'ts to avoid harm and to do good according to conscience in the case of ethics and authority in the case of law.

The two positions, therefore, converge on the same observation from different directions. Antipositivists hold that law ought to rest on moral concerns, and positivists concede that moral and legal norms overlap in content. Neither position requires that a particular ethical text be the origin of a particular legal provision; both describe an association, a shared normative content within a shared context. It is this association, rather than any claim about origins, that the present study treats as observable in the language of the texts themselves.

In philosophical discourse, a single idea can be articulated through a diverse array of linguistic formulations (Rohatyn, 1972), thereby introducing inherent ambiguities that complicate rigorous semantic analysis. This intrinsic variability in the language of philosophy (Adler & Doren, 1972; Gray, 2012; Martinich, 2016) implies that each expression, while conveying a core concept, may simultaneously introduce subtle shifts in meaning or emphasis. Consequently, using NLP techniques to analyze semantic relationships among philosophical texts, especially lengthy ones, creates considerable challenges (Jurafsky & Martin, 2025). To address such analytical difficulties and minimize their impact on the accuracy of meaning, the application of preprocessing techniques becomes a crucial methodological step.

### 2.2. Normative Theories, Law, and the EU AI Act

The three foremost normative theories—namely, virtue ethics, deontological ethics, and consequentialism—do not relate to law in the same way. Each reaches law by a distinct route, and each route operates at a different level of the legal order. This section examines these bridges in turn and the way legal scholarship has read the EU AI Act through each of them.

Deontological ethics has the most direct bridge to law. In the *Doctrine of Right*, Kant (1797/1996) defines an action as right if it can coexist with the freedom of everyone in accordance with a universal law, so that rights and their correlative duties form the content of a just legal order; Ripstein (2009) develops this into a Kantian account of legal and political institutions. Human dignity, whose modern meaning owes much to the Kantian idea of persons as ends in themselves, became a foundational concept of post-war human rights instruments (McCrudden, 2008). It opens the Charter of Fundamental Rights of the European Union: "Human dignity is inviolable. It must be respected and protected" (Charter of Fundamental Rights of the European Union, 2012, Art. 1). Scholarship on the AI Act reads its aims through this bridge. The *Ethics Guidelines for Trustworthy AI* of the High-Level Expert Group on Artificial Intelligence (2019) grounded trustworthy AI in the fundamental rights of the Charter. They derived four ethical principles from them: respect for human autonomy, prevention of harm, fairness, and explicability. Smuha (2019) characterized this as a distinctly European, human-centric approach that anchors AI ethics in fundamental rights, and Ho-Dac (2025) argues that the Act carries this orientation into law, producing a human rights risk-based approach that translates European public interests into the language of fundamental rights and values. The categorical prohibition of practices that pose an unacceptable risk, irrespective of their potential benefits, is the clearest duty-based element of the Act. More broadly, Hagendorff (2020) observes that AI ethics guidelines typically take a deontological form, positing a fixed set of universal principles that developers are expected to adhere to.

Consequentialism reaches law through the design of legislation. Bentham (1789/1996) proposed the principle of utility as the standard by which legislators should judge laws, and the law-and-economics tradition continues this line by evaluating legal rules according to their effects on individual welfare (Kaplow & Shavell, 2002). In EU law-making, this bridge is institutionalized in impact assessment. Before proposing the AI Act, the European Commission (2021) compared alternative policy options by their expected costs and benefits and selected a horizontal instrument with mandatory requirements confined to high-risk applications. The resulting risk-based architecture, borrowed from EU product-safety legislation (Veale & Borgesius, 2021), is where scholarship locates the Act's consequentialist side. Kaminski (2023) argues that regulating AI through risk accepts the deployment of the technology and manages its harms in the aggregate, an approach oriented toward systemic mitigation rather than toward redress for the individuals whose rights are infringed, and Mahler (2022) locates the Act's risk-based approach between risk management and the legal principle of proportionality. Critics contend that this mechanism dilutes the Act's rights-based aims. Laux et al. (2024) argue that the Act conflates the trustworthiness of AI with the acceptability of its risks, Kusche (2024) examines the limits of expressing harms to fundamental rights in the vocabulary of risk, and Rangone and Megale (2025) ask whether the Act regulates risks without adequately protecting rights.

Virtue ethics reaches law through its application rather than its text. Virtue jurisprudence locates the ethical content of law in the character and practical wisdom (*phronēsis*) of those who apply it, above all judges (Solum, 2003). In the same vein, Gürler (2025) argues that the values of judicial ethics, such as independence, impartiality, and integrity, express not only character virtues sought in judges but also the intellectual virtues required for sound legal reasoning. Correspondingly, virtue ethics is largely absent from readings of the AI Act. Hagendorff (2020) argues that the prevailing deontological approach to AI ethics should be augmented with one oriented toward virtue ethics, because principles alone do not address the character and dispositions of those who design and deploy AI systems.

The three bridges thus operate at different levels of law: deontological ethics at the level of legislative content, where rights, duties, and prohibitions are stated; consequentialism at the level of legislative design, where options are weighed by their effects; and virtue ethics at the level of application, where rules are interpreted and applied by people. The literature on the AI Act accordingly describes a hybrid: rights-based, deontological aims implemented through a risk-based, consequentialist mechanism, with little room for virtue ethics in the text itself. These readings are qualitative and interpretive, and they concern the structure and stated purposes of the Act. The present study examines them independently by measuring the textual association between the Act and the three normative theories.

Before delving into the methodology, however, a closer look at the theoretical foundations of semantic association between texts is necessary to develop a comprehensive conceptual framework for this study. Furthermore, it is crucial to separate the relationship between shared meaning and textual similarity from claims about how the texts came to share that meaning. Such an understanding will provide clarity for subsequent analysis and interpretation.

## 2.3. Semantic Association Between Texts

Two texts are semantically associated when they express overlapping meaning: they address related subject matter, draw on related concepts, and use language in related ways. Association is a property of the texts, not of their histories. Two texts can be strongly associated because one drew on the other, because both drew on a common tradition, or because they address the same problem independently; the association itself does not distinguish among these explanations.

Association between texts is a semantic notion. Semantics is a branch of linguistics that deals with the meaning of linguistic elements (Saeed, 2016; Qamar & Raza, 2024). It is not an exact science: among the problems of semantics are ambiguity and vagueness (Kennedy, 2019), which make it difficult to ascribe precise meaning to those linguistic elements.

Meaning, moreover, is not fixed by words in isolation. Halliday (1978) used social semiotics to view language as a dynamic system in which meaning is constructed through the interaction of social structures and linguistic functions, involving ideational, interpersonal, and textual metafunctions.

In later theoretical treatments, Krauss and Fussell (1996) synthesized communication and cognitive models to emphasize the social and contextual nature of meaning construction. Beltrama (2020) focused on social meaning, applying formal semantics and pragmatics to understand how linguistic forms convey information about users' social identities. Both lines of work imply that texts produced within a shared context, such as the moral and legal discourse on harm and good, can be expected to share meaning to a measurable degree.

The computational treatment of such shared meaning rests on the distributional hypothesis: linguistic items that occur in similar contexts tend to have similar meanings (Harris, 1954; Firth, 1957). Modern text representations operationalize this hypothesis by learning vectors from the contexts in which words and sentences occur, so that texts expressing related meaning are mapped to nearby points in a vector space (Jurafsky & Martin, 2025). The proximity of two such representations is therefore a direct, if partial, measure of the association between the texts they represent.

The link between shared meaning and shared representation suggests that the degree of semantic similarity between textual expressions can be used to estimate the extent of their association. STS, a metric used to quantify the degree of semantic equivalence between a pair of texts (Bali et al., 2024), can therefore serve as an indicator of the association between a set of source documents and a target document.

## 2.4. Semantic Textual Similarity

Despite the difficulty of assigning precise meaning to linguistic elements due to ambiguity and vagueness, measuring semantic similarity between two texts remains a fundamental task in NLP (Zhao et al., 2024). By considering the lexical, syntactic, and semantic features of the texts, STS aims to quantify their similarity. Different authors have classified STS methods in different ways (Han et al., 2020; Wang & Dong, 2020; Chandrasekaran & Mago, 2021; He et al., 2024; Sasoko et al., 2024). Based on a combination of Han et al.'s (2020) and He et al.'s (2024) classifications, STS methods can generally be grouped into four major categories: String-based methods, corpus-based methods, knowledge-based methods, and deep-learning methods.

*String-based methods* focus on superficial features of two documents, such as word overlap, n-grams, or string matching. These methods take characters or words from both texts and return an STS score. They work well for duplicate detection and plagiarism analysis, but cannot handle synonyms, paraphrasing, or context shifts. They are computationally efficient but ignore the deeper semantic meaning of the texts. Despite their limitations, string-based approaches are valuable for baseline comparisons and are used during preprocessing stages in NLP pipelines.

*Corpus-based methods* utilize extensive collections of documents, known as corpora, to identify semantic relationships between word pairs. They aim to map words as vectors (word embeddings) in a high-dimensional space, positioning semantically similar words closer together. The STS score between a pair of texts is then derived from the similarity of their respective word embeddings. These methods rely on pre-trained embeddings, such as Word2Vec and GloVe, and often employ similarity metrics, including cosine similarity and Euclidean distance, to quantify semantic closeness. Thus, analyzing linguistic patterns across large datasets helps capture nuances in semantic relationships beyond lexical similarity.

*Knowledge-based methods* use predefined relationships between words and concepts to assess semantic similarity. They use structured linguistic resources, such as ontologies, lexical databases, and semantic networks, to establish semantic relationships between words. Then, they aggregate these similarities to yield an STS score. These methods are particularly valuable because they provide interpretability and domain specificity, which are often lacking in corpus-based models due to data sparsity. However, they struggle with scalability and lack the flexibility to adapt to new linguistic variations beyond predefined taxonomies.

*Deep learning methods* use neural networks to learn semantic representations of text. Unlike corpus-based or knowledge-based approaches, these methods rely on hierarchical feature extraction and contextual embeddings from deep architectures such as Transformers and recurrent networks. They can effectively capture contextual and semantic relationships and directly model the similarity between sentences or passages. Deep-learning STS approaches surpass traditional similarity metrics because they can capture nuanced language relationships, including paraphrasing and implicit meanings. However, they require substantial computational resources and large-volume corpora annotated for effective generalization.

The specific method for obtaining an STS score depends on two main factors: the texts to be compared and the available resources. These two factors can be further refined across the categories of STS methods.

String-based models and corpus-based approaches are transparent and fast but may not capture semantic nuances. A trade-off between explainability and semantic depth is necessary in selecting an STS method (Manning & Schütze, 1999). Corpus and knowledge-based STS methods are suitable in low-resource environments. However, it is essential to balance semantic richness, data scarcity, and computational constraints (Mihalcea et al., 2006).

Agirre et al. (2012) introduced two additional factors: generalizability and human judgment. Generalizability refers to an STS method's ability to maintain reliable performance across different types of texts, domains, and languages. According to the authors, an STS method should consistently yield STS scores that align with human intuition, regardless of the domain in which it is applied. Cer et al. (2017) carried these criteria over to deep-learning methods: they extended STS evaluation to multilingual and cross-lingual settings and introduced the STS Benchmark, a selection of English datasets from earlier STS tasks covering three genres: news (including headlines), image captions, and online forums (including question-answer pairs).

In recent years, deep learning models have taken the lead in STS research, shifting the focus to selecting and applying the most effective methods. Devlin et al. (2019) presented Bidirectional Encoder Representations from Transformers (BERT) to enhance language understanding. Vaswani et al. (2017) proposed the Transformer architecture, which is based on self-attention and served as a precursor to BERT. Self-attention is a mechanism where each element in a word sequence computes a weighted sum of all elements in that same sequence. Learned similarity scores between elements determine their weights. Employing self-attention mechanisms enables parallel computations. Thus, self-attention significantly improves efficiency in NLP tasks that require deep context understanding.

BERT leverages deep bidirectional attention by simultaneously considering both the left and right textual contexts of a string. It enhances NLP task performance by employing a pre-training strategy that combines Masked Language Modeling (MLM) and Next Sentence Prediction (NSP). MLM is based on random masking and word prediction. NSP helps the model identify relationships between sentences. These pre-training strategies enable more precise semantic representations and improve contextual understanding. With minimal modifications, developers fine-tuned BERT for tasks such as text classification, named entity recognition, and question answering.

Reimers and Gurevych (2019) adapted BERT to Sentence-BERT (SBERT), which uses Siamese and triplet network structures to encode each sentence into a fixed-size embedding independently. Sentences can then be compared using cosine similarity rather than being passed through the network in pairs, reducing the cost of large-scale similarity search from hours to seconds. The SBERT checkpoint used in this study, all-MPNet-base-v2, applies this framework to MPNet (Song et al., 2020), an encoder pre-trained with combined masked and permuted language modeling, and fine-tunes it with a contrastive objective on a large collection of sentence pairs.

Similarly, Lan et al. (2020) proposed A Lite BERT (ALBERT) to address the limitations of the original BERT model without compromising performance. The authors employed two parameter-reduction techniques, factorized embedding parameterization and cross-layer parameter sharing, to reduce resource requirements. They also replaced BERT's next-sentence prediction with a sentence-order prediction loss, which focuses on inter-sentence coherence and improves performance on tasks with multi-sentence inputs. The checkpoint used in this study, paraphrase-albert-small-v2, is a small ALBERT variant trained for sentence embedding on paraphrase data.

Another low-compute Transformer model, DistilBERT, was presented by Sanh et al. (2019). The model is based on knowledge distillation to develop a smaller general-purpose language representation. Their approach can be fine-tuned for various tasks with only a slight performance sacrifice: it is 40% smaller and 60% faster than BERT, while retaining 97% of its language-understanding performance. The checkpoint used in this study, distilbert-base-nli-stsb-mean-tokens, was fine-tuned on natural language inference and STS benchmark data, with mean pooling over token embeddings. Its developers have since deprecated this checkpoint for producing lower-quality embeddings than newer models; it is retained here for continuity with earlier versions of this study and with Albayrakoglu and Aydin (2026).

Liu et al. (2019) introduced an improved version of BERT, known as the Robustly Optimized BERT Pretraining Approach (RoBERTa). The authors removed the Next Sentence Prediction (NSP) from the original model. They trained it on longer sequences than those used in BERT and employed dynamic masking to enhance their model's accuracy and performance. Dynamic masking enables the model to learn from multiple masking patterns per sentence, thereby adapting to various sentence structures. The checkpoint used in this study, all-distilroberta-v1, is a distilled six-layer RoBERTa fine-tuned contrastively for sentence embedding.

Jiao et al. (2020) introduced TinyBERT, a compact and efficient variant of BERT designed to reduce computational requirements while preserving performance. TinyBERT employs a two-stage knowledge distillation framework, comprising pretraining and fine-tuning. As a result, the model is substantially smaller and faster than BERT while achieving comparable performance on NLP tasks. The checkpoint used in this study, paraphrase-TinyBERT-L6-v2, is a six-layer TinyBERT trained for sentence embedding on paraphrase data.

Wang et al. (2020) proposed MiniLM, which compresses a large Transformer through deep self-attention distillation: the student model learns to mimic the self-attention distributions and value relations of the teacher's last Transformer layer, rather than its layer-by-layer outputs. This task-agnostic approach yields a small, fast model that retains most of the teacher's accuracy. The checkpoint used in this study, all-MiniLM-L12-v2, is a 12-layer MiniLM with 384-dimensional embeddings, fine-tuned contrastively on a large collection of sentence pairs.

Wang et al. (2024) introduced E5, a family of general-purpose text embedding models trained with weakly supervised contrastive pre-training on a large, curated corpus of text pairs mined from the web, followed by supervised fine-tuning on labeled data. E5 expects each input to carry a prefix that signals its role; following the developers' recommendation for symmetric tasks such as STS, every sentence in this study was prefixed with "query: ". The checkpoint used in this study is e5-base-v2.

Finally, Li et al. (2023) proposed General Text Embeddings (GTE), trained with multi-stage contrastive learning: unsupervised pre-training on a large mixture of weakly related text pairs from diverse sources, followed by supervised fine-tuning on multiple annotated datasets. Despite its moderate size, GTE-base performs competitively with considerably larger embedding models. The checkpoint used in this study is gte-base.

While presenting their models, the developers of the eight encoders also mentioned their criteria for selecting the most suitable model, either explicitly or implicitly. Table 1 summarizes the checkpoints used in this study, their main characteristics, and the criteria their developers stated or implied.

This study employs these eight Transformer-based sentence encoders, SBERT, ALBERT, DistilBERT, RoBERTa, TinyBERT, MiniLM, E5-base, and GTE-base, to compute semantic textual similarity (STS). BERT itself is not used directly; it is described above as the architecture from which most of the encoders descend. The following section details the research design, preprocessing procedures, and model application steps used to operationalize this approach.

**Table 1** Transformer-based sentence encoders used in this study: checkpoints, characteristics, and dominant selection criteria

| Encoder | Checkpoint | Training objective | Param. | Dim. | Max. seq. | Dominant selection criteria |
|---|---|---|---|---|---|---|
| *SBERT* | all-MPNet-base-v2 | Contrastive learning | ~110M | 768 | 384 | Sentence-level semantics; low-resource and real-time suitability |
| *ALBERT* | paraphrase-albert-small-v2 | Paraphrase training | ~12M | 768 | 100 | Memory and parameter efficiency; scalable similarity scoring |
| *DistilBERT* | distilbert-base-nli-stsb-mean-tokens | NLI + STS fine-tuning | ~66M | 768 | 128 | Reduced model size with acceptable accuracy |
| *RoBERTa* | all-distilroberta-v1 | Contrastive learning | ~82M | 768 | 512 | Performance optimization within existing criteria; accuracy maximization through improved training |
| *TinyBERT* | paraphrase-TinyBERT-L6-v2 | Paraphrase training | ~67M | 768 | 128 | Extreme compression; accuracy-cost trade-offs |
| *MiniLM* | all-MiniLM-L12-v2 | Contrastive learning | ~33M | 384 | 128 | Task-agnostic compression; fast inference with retained accuracy |
| *E5-base* | intfloat/e5-base-v2 | Contrastive pre-training | ~110M | 768 | 512 | General-purpose embeddings transferable across tasks without task-specific tuning |
| *GTE-base* | thenlper/gte-base | Multi-stage contrastive | ~110M | 768 | 512 | General-purpose embeddings from diverse training data; strong performance at moderate size |

Param. = total learned weights; Dim. = dimensionality of the sentence embedding; Max. seq. = maximum input length in tokens. Sources for each encoder are given in the text.

# 3. Methodology

Figure 1 summarizes the study's research design and analytical workflow. Three canonically distinct normative theories of ethics are selected as source documents. At the same time, the EU AI Act, the target document, is partitioned into two target sections, its preamble and statutory provisions, to distinguish its intentional and operational aspects. Following high-level text preprocessing of the source documents to minimize semantic overlap, every document is segmented into sentences, and each sentence is encoded by eight Transformer-based sentence encoders. Each source document is compared with each target section, and the three source documents are also compared with one another. Sentence embeddings are aggregated at the document level using three methods, and the resulting scores are analyzed for ranking agreement across encoders, target sections, and levels of granularity.

## 3.1. The Data: Three Normative Theories of Ethics and the EU AI Act

This subsection summarizes the source documents, encyclopedic treatments of virtue ethics, deontological ethics, and consequentialism, and the target document, the EU AI Act. Although scholars wrote them, there are several reasons for choosing encyclopedic entries as source documents over original philosophical treatments, scholarly books, articles, or textbook narratives. First, encyclopedia entries help avoid subjective or argumentative narratives, which are likely to complicate the machine's understanding of texts. Second, the materials should not be targeted at a specific segment of the audience. Third, the degree of textual structure of the source and target documents should be as compatible as possible. Fourth, a consistent, up-to-date terminology and vocabulary should be used in narratives drawn from a range of philosophical resources, including translations and historical works.

The first normative theory, *virtue ethics* (Hursthouse & Pettigrove, 2023), emphasizes the importance of virtues and a person's moral character in action. The ethical decisions and actions of a virtuous person, who strives to do what is right, good, just, or proper, are based on their character and conduct. Virtue ethics is based on three ideas from ancient Greek philosophy: *phronesis* (moral or practical wisdom), *eudaemonia* (happiness or flourishing), and *arete* (excellence or virtue). Virtues are admirable character traits that guide a person's attitudes and actions. An ethical person is honest, wise, fair, courageous, and self-controlled. However, virtue is a matter of degree; perfect or flawless virtue is uncommon.

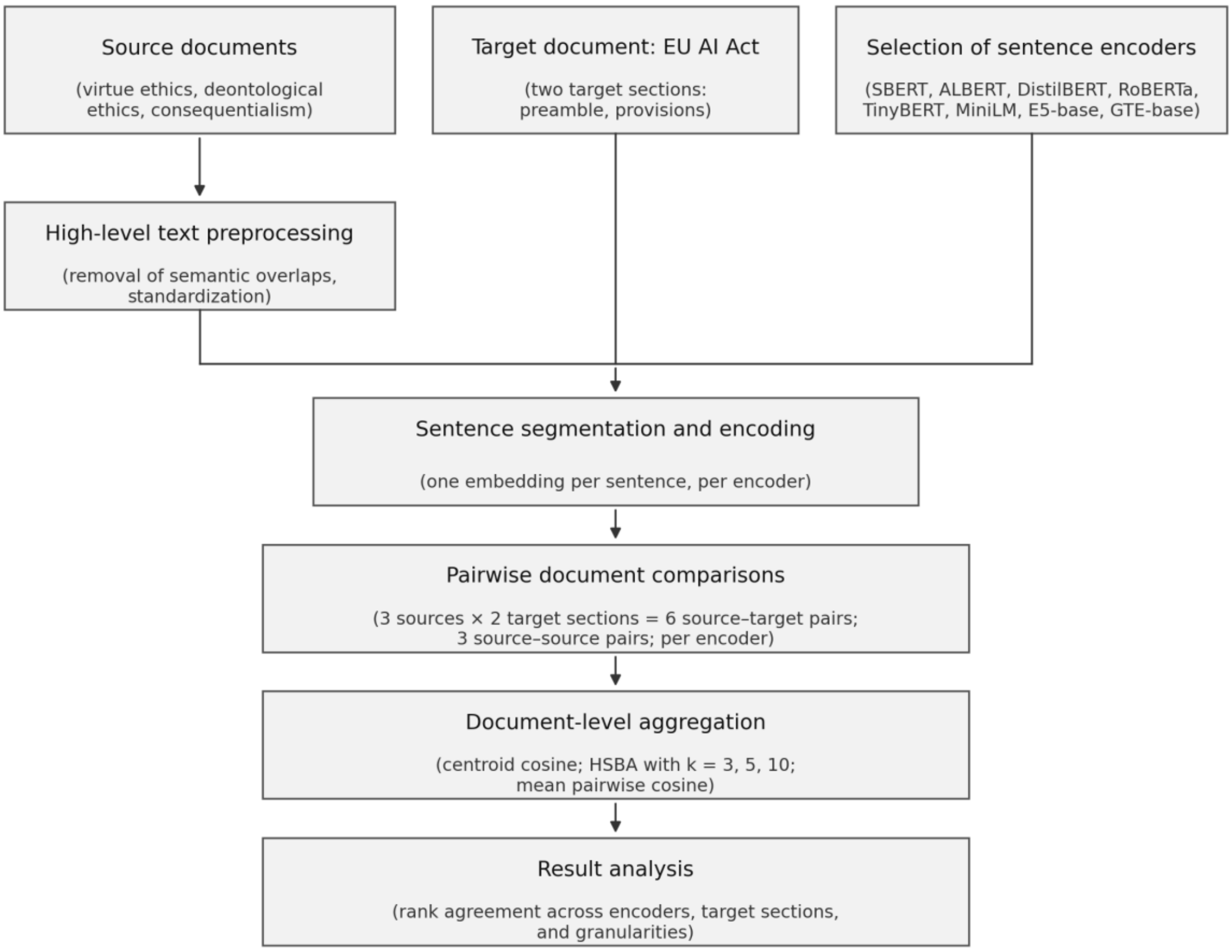

**Figure 1** Flowchart of the research design and analytical workflow of the study

The second normative theory, *deontological ethics* (Alexander & Moore, 2021), categorizes actions as morally required, forbidden, or permitted. It guides and assesses a person's choices of what they ought to do. Deontological approaches hold that some options are morally forbidden even if their overall effect would be good. Deontologists believe that a choice is right if it conforms to a moral norm that each moral agent should obey. The best-known form of deontological ethics comes from Kant, whose central principle, the categorical imperative, is usually stated in three ways. First, before acting, ask whether the reason for your action could serve as a rule that everyone follows. Second, treat every person, including yourself, as valuable in their own right, never merely as a means to an end. Third, act as if you were helping to make the rules for a community in which everyone is both a rule-maker and bound by the same rules. Some deontologists focus on agency and the idea that morality is, to some extent, a personal matter.

The final normative theory, *consequentialism* (Sinnott-Armstrong, 2023), suggests that moral rightness or wrongness depends solely on the consequences of one's decisions and actions. *Hedonism* holds that pleasure is the only intrinsic good and pain the only intrinsic evil. Classic utilitarians hold a hedonistic act-consequentialist view, which claims that an act is morally justified if it causes the greatest happiness for the greatest number of stakeholders. Additional normative characteristics included in consequentialist theories should depend solely on consequences. There are several shades of consequentialist theories, such as maximizing, hedonistic, and aggregative consequentialism. What distinguishes one view from another is the extent to which moral rules are included.

These theories have some commonalities. First, any description of a theory of ethics should be centered around ethically acceptable or unacceptable attitudes, decisions, or acts. Ethical behavior lies on a continuum between totally acceptable or positive, and unacceptable or negative. On the positive end, human acts are often qualified as right, good, fair, virtuous, appropriate, beneficial, impartial, and unbiased, among other terms. On the negative end, they are wrong, bad, unfair, vicious, inappropriate, harmful, partial, biased, etc.

Apart from this central theme, there are other similar aspects in theories of ethics. Table 2 qualitatively lists the most significant similarities between each pair of the three major normative theories of ethics, derived from pairwise comparisons of the preprocessed entries from the Stanford Encyclopedia of Philosophy (Alexander & Moore, 2021; Hursthouse & Pettigrove, 2023; Sinnott-Armstrong, 2023), and the works of Kagan (1998) and Wood (2020).

On the table, all three theories claim to be normative and universal, while also taking circumstances into account and acknowledging multiple perspectives on a good life. Additionally, each theory rejects pure forms of the others while maintaining normativity and universality, suggesting that these frameworks are complementary rather than mutually exclusive.

There are also pairwise similarities between the theories. Virtue and deontological ethics emphasize practical reason, moral excellence, character, duty-based thinking, and the priority of right over good, each to a significant, yet varying degree. Deontological ethics and consequentialism share an emphasis on moral reasoning, impartiality, stakeholder relations, consideration of circumstances, the importance of intention, and the priority of good over mere praiseworthiness. Finally, virtue ethics and consequentialism share multiple views of a good life, the presence of idealized moral agents, consideration of circumstances, and the importance of happiness and human relationships.

**Table 2** A qualitative summary of similarities between pairs of the three normative theories of ethics

| **Virtue Ethics and Deontological Ethics** | **Virtue Ethics and Consequentialism** | **Deontological Ethics and Consequentialism** |
|---|---|---|
| Normativeness | Normativeness | Normativeness |
| Universality | Universality | Universality |
| Priority of right over good | Multiple views of a good life | Multiple views of a good life |
| Emphasis on moral excellence | Presence of idealized moral agents | Emphasis on moral reasoning |
| Importance of practical reason | Consideration of circumstances | Consideration of circumstances |
| Importance of obligations | Importance of happiness | Emphasis on impartiality |
| Importance of character | Importance of human relationships | Emphasis on stakeholder relations |
| Importance of intention | Rejection of pure deontology | Priority of good over praiseworthy |
| Rejection of pure utility | Rejection of pure emotions | Importance of intention |
| | Rejection of pure rationality | |

The EU AI Act (European Parliament & Council of the European Union, 2024) regulates the development and use of AI systems across EU member states. It also applies to non-EU companies operating inside the EU. The Act's objectives are to ensure the safe use of AI systems, support fundamental rights, and foster AI innovation within the EU.

The Act defines four categories of risk for AI systems. The highest level is *unacceptable risk*, which refers to AI systems and practices regarded as harmful or unethical. Such systems threaten fundamental EU values and rights, and, consequently, the Union prohibits their use across its member states. *High-risk* systems are those used in mission-critical sectors, such as healthcare and law enforcement, or those that may compromise fundamental rights. These systems are subject to rigorous controls and oversight. *Limited-risk* systems can be used for deception and manipulation and require specific transparency measures. *Minimal-risk* systems pose almost no risk to individuals' safety or fundamental rights. They are not subject to any particular regulatory obligation.

In addition to risk categories, the EU AI Act emphasizes the fundamental rights of EU citizens, including human dignity, freedom, equality, and democracy. It also encourages the rule of law and respect for human rights. The Act distinguishes between the developers and users of AI systems and specifies several obligations for both groups for high-risk systems.

Due to the increasing capabilities and potential impact of recent AI systems, such as Generative AI (GenAI) applications, the Act introduces specific transparency requirements for their developers, regardless of their intended purpose of use. As these applications become more powerful, they are subject to additional, stricter requirements for model evaluation, risk assessment and mitigation, incident reporting, and cybersecurity.

The law also introduces a robust governance and enforcement framework, including the establishment of an EU AI Office within the European Commission (EC) and the requirement that member states create national AI offices. It sets significant penalties for non-compliance, depending on the severity of the infringement and the size of the EU and non-EU developers or users of the systems. Finally, the Act establishes provisions for scope changes, recognizing the dynamic nature of AI technology.

For the analysis, the Act is divided into its two target sections: the preamble, comprising the 180 recitals that state the Act's reasons and purposes, and the statutory provisions, comprising its enacting terms. After sentence segmentation, the preamble contains 883 sentences and the provisions 1,652 sentences; the preprocessed source documents contain 148 (virtue ethics), 177 (deontological ethics), and 280 (consequentialism) sentences.

## 3.2. Justification of the Central Proposition

The literature review has established that semantic similarity can serve as an indicator of association between documents that share a context. The discussion of the ethics of law has further demonstrated that ethics and law share such a context: antipositivists and some positivists recognize that moral and legal norms overlap in content and purpose. The condition of the central proposition is therefore satisfied for the three normative theories and the EU AI Act.

Although association does not require temporal precedence, the chronology of the source and target documents helps characterize their relationship. It is established by comparing the publication years of each major official EU AI Act document with those of the resources consulted in the bibliography of each source document.

The first initiative leading to the EU AI Act started with the Consultation on Artificial Intelligence, launched in February 2020. The results were published in a white paper titled "Public Consultation on the AI White Paper: Final Report" in November 2020 (Directorate-General for Communications Networks, Content and Technology, 2020). In April 2021, the proposal was presented to the European Parliament by the Commission's Directorate-General for Communications Networks, Content, and Technology (2021). Finally, it was enacted by the European Parliament in March 2024, approved by the European Union Council in May 2024, and came into force on August 1, 2024, with some provisions covering up to 3 years after the enforcement of the law (European Parliament & Council of the European Union, 2024).

The descriptions of the theories of ethics are taken from the Stanford Encyclopedia of Philosophy (Zalta & Nodelman, n.d.). The bibliography of the entry Virtue Ethics spans the period between 1956 and 2021 (Hursthouse & Pettigrove, 2023), although its roots can be traced back to Plato (429 BC-347 BC) and Aristotle (384 BC-322 BC) (Van Zyl, 2019). The bibliography of Deontological Ethics spans the 18th century to 2019 (Alexander & Moore, 2021). Likewise, Consequentialism's bibliography starts in 1755 and ends in 2020 (Sinnott-Armstrong, 2023).

Compared to the period covered by the EU AI Act documentation, the bibliographies of the normative theories of ethics originated from works written centuries earlier. The three source documents thus draw on traditions that predate the target document. This chronology is consistent with, but not evidence of, a directional relationship; the analysis therefore restricts its claims to association.

## 3.3. Semantic Textual Similarity (STS) with Transformer-Based Sentence Encoders

This study uses the eight encoders discussed in the literature review to calculate STS. Semantic space is considered a normalized metric space in which distance is used to measure semantic similarity (Rozinek & Mareš, 2021). Distinct from the lexical semantics that apply to words, STS applies to larger units of language: sentences, paragraphs, or longer pieces consisting of multiple paragraphs, sections, chapters, parts, or entire textual artifacts. The term "metric" refers to the measurement of the distance between two points in space (Dshalalow, 2013).

Transformer-based sentence encoders are compared using *cosine* similarity, which emphasizes direction over magnitude. Since it does not directly satisfy the metric space axioms, cosine similarity is referred to as a *pseudo-metric*. Nevertheless, it can be converted to a cosine distance through an $arccosine$ transformation, and cosine distance is a metric.

In the classical vector space model, a term-frequency vector, consisting of the number of occurrences of each term in a document, represents the document. Similarity between two documents is calculated by applying the following formula to their vectors:

$$\sigma(\boldsymbol{x}, \boldsymbol{y}) = \frac{\boldsymbol{x} \cdot \boldsymbol{y}}{||\boldsymbol{x}||\ ||\boldsymbol{y}||} \qquad \text{(Equation 1)}$$

where

$\sigma(\boldsymbol{x}, \boldsymbol{y})$ = Similarity of two term-frequency vectors $\boldsymbol{x}$ and $\boldsymbol{y}$

$\boldsymbol{x}$ = term-frequency vector of document $x$

$\boldsymbol{y}$ = term-frequency vector of document $y$

$||\boldsymbol{x}||$ = Euclidean norm of the vector $\boldsymbol{x}$

$||\boldsymbol{y}||$ = Euclidean norm of the vector $\boldsymbol{y}$

$\sigma(\boldsymbol{x}, \boldsymbol{y})$ is a measure of how close two non-zero vectors are in an inner product space. The closer the pair of vectors is, the more similar the two documents (Han et al., 2012). The same formula applies to the dense sentence embeddings produced by Transformer-based encoders, in which case $\boldsymbol{x}$ and $\boldsymbol{y}$ are learned vectors rather than term-frequency vectors.

## 3.4. Text Preprocessing

Text preprocessing is crucial in NLP because raw text often contains noise, inconsistencies, and redundancies that can negatively impact model performance. In Semantic Textual Similarity (STS), preprocessing ensures that models accurately capture meaning rather than surface-level differences. In the STS literature, however, text processing refers to operations such as tokenization, lowercasing, stopword removal, and stemming applied to the lexical elements of a text (Chai, 2023).

To avoid confusion about what each source document is about, the higher-level preprocessing rules shown in Table 3 were uniformly applied to the sentences, paragraphs, and words in the descriptions of the theories of ethics (Albayrakoglu & Aydin, 2026). The purpose of using these rules is to minimize semantic overlap among the descriptions of theories. These rules help eliminate linguistic elements that could interfere with the meaning of descriptions, as they are more relevant to another theory from a machine learning perspective. Additionally, each rule was justified by adding a rationale immediately following it. In applying these rules, care was taken to ensure that no rule altered evaluative content, normative claims, or core vocabulary.

Once the manual preprocessing of the theories was complete, the target document was divided into its two target sections, the preamble and the provisions, to examine whether their patterns of association with the theories are consistent. Both target sections were used as-is without modification.

**Table 3** Text preprocessing rules applied to theories of ethics.

| Rule No. | Description | Rationale |
|---|---|---|
| 1 | Remove titles, subtitles, etc. | They do not describe a theory but indicate specific parts of the document. |
| 2 | Eliminate meta descriptions (descriptions of the document, e.g., TOC, abstract, etc.) | They do not describe the theory; instead, they show what the document is about or how it is organized. |
| 3 | Remove items from the reference list. | They do not describe a theory. |
| 4 | Remove proper nouns. | Not the nouns but the ones they belonged to described each theory. |
| 5 | Delete discussions about what the theory is not, but keep negative examples. | Those discussions do not describe a theory. |
| 6 | Keep only the conclusive statements for the incremental arguments. | Eliminate irrelevant words. |
| 7 | Keep comparisons of the various forms of the same theory. | They are indispensable extensions of a theory. |
| 8 | Remove references to religions and religious symbols. | Ensure religious neutrality. |
| 9 | Delete descriptions of, references to, or comparisons with other theories. | Isolate one theory from another to preserve context. |
| 10 | Convert text into US English. | Eliminate variations in spelling. |
| 11 | Replace foreign-language words (mostly Greek or Latin) with their US-English equivalents, if any (Collins and Merriam-Webster). | Reduce the likelihood of encountering missing words in the model. |
| 12 | Add English translations of foreign-language words if there is no US-English equivalent. | Reduce the likelihood of encountering missing words in the model. |

## 3.5. Sentence Encoding

Each document is segmented into an ordered sequence of sentences with NLTK's sent_tokenize function (Bird et al., 2009). Each non-empty line of the text file is treated as a hard boundary within which sentence segmentation is applied, and segments shorter than five characters are discarded. Every sentence is then encoded independently, once per encoder, through the Sentence-Transformers library (Reimers & Gurevych, 2019), with the pooling configuration distributed with each checkpoint. Five checkpoints (SBERT, RoBERTa, MiniLM, E5-base, and GTE-base) return L2-normalized sentence embeddings; the other three (ALBERT, DistilBERT, and TinyBERT) do not. Encoding sentences individually avoids the silent truncation that occurs when a whole document is passed to an encoder with a maximum input length of a few hundred tokens.

Sentences longer than an encoder's maximum sequence length are still truncated, but a token-length audit shows that this affects few sentences. No sentence exceeds the limit of the four encoders that accept 384 or 512 tokens. For the three encoders limited to 128 tokens (DistilBERT, TinyBERT, and MiniLM), 18 of the 883 preamble sentences (2.0%), 6 of the 1,652 provision sentences (0.4%), and 2 of the 177 deontological-ethics sentences (1.1%) are truncated. ALBERT, with a limit of 100 tokens, truncates 53 preamble sentences (6.0%), 29 provision sentences (1.8%), 3 deontological-ethics sentences (1.7%), and 1 consequentialism sentence (0.4%). Truncation removes only the trailing tokens of these sentences.

## 3.6. Document-Level Aggregation

Comparing two documents requires aggregating their sentence embeddings into a single score. Three aggregation methods are used; they differ in the granularity at which the documents are compared. The notation follows Albayrakoglu and Aydin (2026): a document $D$ is an ordered set of $n$ sentences $s_1, \ldots, s_n$, and an encoder $f$ maps each sentence $s_i$ to an embedding $\boldsymbol{e}_i \in \mathbb{R}^d$.

*Centroid cosine similarity* compares documents as wholes. The centroid of a document is the arithmetic mean of its sentence embeddings, and the similarity of two documents $A$ and $B$ is the cosine of the angle between their centroids:

$$\boldsymbol{c}_D = \frac{1}{n}\sum_{i=1}^{n} \boldsymbol{e}_i \qquad \text{(Equation 2)}$$

$$S_C(A,B) = \frac{\boldsymbol{c}_A \cdot \boldsymbol{c}_B}{\|\boldsymbol{c}_A\|\,\|\boldsymbol{c}_B\|} \qquad \text{(Equation 3)}$$

The centroid represents a document's overall semantic orientation. Because it is a mean, it is not affected by document length, and a sentence that is unrelated to the other document shifts it only in proportion to its share of the document. Centroid similarity is the primary measure of this study.

*Mean pairwise sentence similarity* compares documents sentence by sentence. It is the arithmetic mean of the cosine similarities of all $m \times n$ cross-document sentence pairs, where $m$ and $n$ are the numbers of sentences in documents $A$ and $B$, and $\mathbf{e}_i^A$ and $\mathbf{e}_j^B$ are their sentence embeddings:

$$S_P(A,B) = \frac{1}{mn}\sum_{i=1}^{m}\sum_{j=1}^{n} cos\left(\boldsymbol{e}_i^A, \boldsymbol{e}_j^B\right) \qquad \text{(Equation 4)}$$

When two long documents address many subtopics, most sentence pairs are unrelated, and their low similarities dominate the mean (Albayrakoglu & Aydin, 2026). The measure has a further, more fundamental property. Let $\boldsymbol{u}_i = \boldsymbol{e}_i / \|\boldsymbol{e}_i\|$ denote the unit-length embedding of the sentence $i$, let $\bar{\boldsymbol{u}}_D$ be the mean of the unit-length embeddings of the document $D$, and let $R_D = \|\bar{\boldsymbol{u}}_D\|$. Because the mean of the pairwise inner products of unit vectors equals the inner product of their means,

$$S_P(A,B) = \bar{\boldsymbol{u}}_A \cdot \bar{\boldsymbol{u}}_B = R_A R_B cos(\bar{\boldsymbol{u}}_A, \bar{\boldsymbol{u}}_B) \qquad \text{(Equation 5)}$$

In directional statistics, $R_D$ is known as the mean resultant length (Mardia & Jupp, 2000). It lies between 0 and 1 and measures how tightly the sentences in a document cluster around a common direction: it approaches 1 when all sentences point in the same direction and falls toward 0 as they scatter. For encoders that return unit-length embeddings, $\bar{\boldsymbol{u}}_D$ coincides with the centroid $\boldsymbol{c}_D$, so $S_P(A,B) = R_A R_B S_C(A,B)$. Mean pairwise similarity is therefore the centroid similarity scaled by the internal coherence of the two documents. It carries no information about their association beyond $S_C$, but mixes that association with each document's own property. It is reported for completeness.

*Hierarchical sentence-block aggregation* (HSBA) compares documents at an intermediate granularity. The ordered sentence sequence of a document is partitioned into consecutive, non-overlapping blocks of $k$ sentences, with the final, possibly incomplete block retained, which gives $T_D = \text{ceiling}(n/k)$ blocks for a document $D$ with $n$ sentences. Block $t$ $(t = 1, \ldots, T_D)$ contains the sentences with indices $B_t = \{(t-1)k + 1, \ldots, \min(tk, n)\}$, so $|B_t| = k$ for every block except possibly the last. Each block embedding is the mean of the sentence embeddings in the block, and HSBA is the mean cosine similarity over all cross-document block pairs, where $\mathbf{b}_t^A$ and $\mathbf{b}_{t'}^B$ denote the $t$-th block embedding of the document $A$ and the $t'$-th block embedding of the document $B$:

$$\boldsymbol{b}_t = \frac{1}{|B_t|}\sum_{i \in B_t} \boldsymbol{e}_i \qquad \text{(Equation 6)}$$

$$S_H^{(k)}(A,B) = \frac{1}{T_A T_B} \sum_{t=1}^{T_A} \sum_{t'=1}^{T_B} cos\,(\boldsymbol{b}_t^A, \boldsymbol{b}_{t'}^B) \qquad \text{(Equation 7)}$$

By the same argument as Equation 5,

$$S_H^{(k)}(A,B) = R_A^{(k)} R_B^{(k)} cos\left(\bar{\boldsymbol{v}}_A^{(k)}, \bar{\boldsymbol{v}}_B^{(k)}\right) \qquad \text{(Equation 8)}$$

where $\bar{\boldsymbol{v}}_D^{(k)}$ is the mean of the unit-length block embeddings of the document $D$ and $R_D^{(k)}$ is its length. HSBA thus spans the range between the other two measures. When $k = 1$, each block is a single sentence, and HSBA equals the mean pairwise similarity. When $k \geq n$, each document forms a single block, and HSBA equals centroid similarity.

As $k$ grows, averaging within blocks cancels part of the scatter among sentences, so $R_D^{(k)}$ rises toward 1, and HSBA rises toward the centroid similarity. This study evaluates $k = 3, 5,$ and 10 as a sensitivity analysis over granularity.

A block-level variant that replaces pairwise comparisons with the cosine of block centroids would add nothing. Because averaging is associative, the mean of the block means equals the document centroid, except for the small extra weight given to the sentences in a final incomplete block. It is the normalization of each block before the pairwise comparison that distinguishes HSBA from centroid similarity.

### 3.7. Multi-Encoder Rank-Voting Ensemble

In NLP, ensemble approaches can combine representations or outputs from multiple models to improve robustness in semantic similarity analysis. All eight encoders used in this research are sentence-embedding models: each has been fine-tuned so that the pooled output of its Transformer layers represents the meaning of a whole sentence.

The application of ensemble methods in text mining involves using different algorithms and models to validate results (Zong et al., 2021). We employed a multi-encoder rank-voting approach to balance semantic expressiveness with computational efficiency and methodological reproducibility, and to reduce model-specific biases (Dietterich, 2000; Zhou, 2012).

The encoders are combined by voting on rankings rather than by averaging raw scores. Transformer embedding spaces have different cosine-similarity baselines and are anisotropic to different degrees (Ethayarajh, 2019): in this study, E5-base and GTE-base return centroid similarities between 0.92 and 0.95 for every source-target pair, whereas the other six encoders spread their six source-target scores over ranges of 0.12 to 0.20. An average of raw scores across encoders would therefore be driven largely by the encoders with the highest baselines. Raw scores are compared only within an encoder, and each encoder yields one ranking of the source documents for each target section and each aggregation method.

## 4. Results

Once the text preprocessing was complete, each source document was compared with each target section of the Act under each of the eight encoders and the three aggregation methods. Table 4 reports the centroid cosine similarities for the preamble and the provisions. The column captioned "Checkpoint" identifies the pre-trained model used for each encoder. In both target sections, all eight encoders rank the three source documents in the same order: deontological ethics first, consequentialism second, and virtue ethics third. Deontological ethics thus receives all 16 votes for the highest association, and the full ranking is unanimous.

The margins differ across encoders in absolute terms but are consistent within them. Deontological ethics leads the second-ranked source document by 0.060 to 0.109 in SBERT, ALBERT, RoBERTa, and MiniLM, and by 0.025 to 0.067 in DistilBERT and TinyBERT. In E5-base and GTE-base, the lead is only 0.011 to 0.015, but it is large relative to the band of about 0.02 within which these two encoders place all six source-target scores. The ordering of consequentialism above virtue ethics is less decisive: the gap is below 0.005 across the SBERT provisions and in both target sections for E5-base and GTE-base.

A second pattern concerns the target sections. For every encoder and every source document, the preamble is more closely associated with the source document than the provisions are. The difference is largest for DistilBERT (0.137 on average) and TinyBERT (0.057), between 0.024 and 0.031 for SBERT, ALBERT, RoBERTa, and MiniLM, and 0.005 to 0.006 for E5-base and GTE-base. The pattern holds without exception under all three aggregation methods and at every block size.

Figure 2 shows how the scores change with the granularity of aggregation. For each encoder, it traces the three source documents from mean pairwise similarity ($k$ = 1) through HSBA with $k$ = 3, 5, and 10 to centroid similarity ($k = n$). In all 48 encoder-source-section combinations, the three HSBA scores lie between the mean pairwise and centroid scores and increase with $k$, consistent with Equation 8. Deontological ethics remains first at every granularity, for every encoder and in both target sections. The positions of the other two source documents, however, depend on granularity. At the document level, consequentialism is second in all 16 encoder-section combinations, but at finer granularity, virtue ethics overtakes it in some encoders: in 6 of the 16 combinations under mean pairwise similarity, 8 under HSBA with $k$ = 3, and 7 under both $k$ = 5 and $k$ = 10 (Table 5).

**Table 4** Centroid cosine similarity between each source document and the two target sections of the EU AI Act

| | | Preamble | | | Provisions | | |
|---|---|---|---|---|---|---|---|
| **Encoder** | **Checkpoint** | **VE** | **DE** | **CO** | **VE** | **DE** | **CO** |
| *SBERT* | all-MPNet-base-v2 | 0.351 | **0.440** | 0.361 | 0.323 | **0.432** | 0.323 |
| *ALBERT* | paraphrase-albert-small-v2 | 0.412 | **0.573** | 0.490 | 0.373 | **0.559** | 0.456 |
| *DistilBERT* | distilbert-base-nli-stsb-mean-tokens | 0.596 | **0.655** | 0.630 | 0.460 | **0.539** | 0.472 |
| *RoBERTa* | all-distilroberta-v1 | 0.277 | **0.411** | 0.351 | 0.241 | **0.393** | 0.312 |
| *TinyBERT* | paraphrase-TinyBERT-L6-v2 | 0.445 | **0.543** | 0.517 | 0.381 | **0.502** | 0.450 |
| *MiniLM* | all-MiniLM-L12-v2 | 0.319 | **0.474** | 0.391 | 0.296 | **0.458** | 0.357 |
| *E5-base* | intfloat/e5-base-v2 | 0.926 | **0.940** | 0.927 | 0.918 | **0.936** | 0.921 |
| *GTE-base* | thenlper/gte-base | 0.935 | **0.949** | 0.938 | 0.929 | **0.946** | 0.933 |
| | **Votes for the highest association** | 0 | 8 | 0 | 0 | 8 | 0 |
| | **Mean rank** | 3.00 | 1.00 | 2.00 | 3.00 | 1.00 | 2.00 |

VE = virtue ethics; DE = deontological ethics; CO = consequentialism. The highest score in each encoder and target section is in bold. Raw scores are comparable only within an encoder.

**Table 5** Agreement of the eight encoders on the ranking of the source documents, by aggregation method and target section

| **Aggregation method** | **Target section** | **DE first** | **CO second** | **VE second** | **Mean rank VE** | **Mean rank DE** | **Mean rank CO** |
|---|---|---|---|---|---|---|---|
| Mean pairwise ($k$ = 1) | Preamble | 8/8 | 5/8 | 3/8 | 2.62 | 1.00 | 2.38 |
| | Provisions | 8/8 | 5/8 | 3/8 | 2.62 | 1.00 | 2.38 |
| HSBA, $k$ = 3 | Preamble | 8/8 | 4/8 | 4/8 | 2.50 | 1.00 | 2.50 |
| | Provisions | 8/8 | 4/8 | 4/8 | 2.50 | 1.00 | 2.50 |
| HSBA, $k$ = 5 | Preamble | 8/8 | 5/8 | 3/8 | 2.62 | 1.00 | 2.38 |
| | Provisions | 8/8 | 4/8 | 4/8 | 2.50 | 1.00 | 2.50 |
| HSBA, $k$ = 10 | Preamble | 8/8 | 5/8 | 3/8 | 2.62 | 1.00 | 2.38 |
| | Provisions | 8/8 | 4/8 | 4/8 | 2.50 | 1.00 | 2.50 |
| Centroid ($k = n$) | Preamble | 8/8 | 8/8 | 0/8 | 3.00 | 1.00 | 2.00 |
| | Provisions | 8/8 | 8/8 | 0/8 | 3.00 | 1.00 | 2.00 |

Entries in the first three numeric columns are votes (number of encoders, out of eight). VE = virtue ethics; DE = deontological ethics; CO = consequentialism.

Equation 5 explains this reversal. For unit-normalized sentence embeddings, mean pairwise similarity and HSBA multiply the similarity between the two documents' average meanings (their mean directions) by the coherence of each. The ratio of mean pairwise similarity to centroid similarity confirms this factorization empirically: for the five encoders that return unit-length embeddings, the ratios for all nine source-target and source-source comparisons decompose into one coherence value per document to within $10^{-5}$; for ALBERT, DistilBERT, and TinyBERT, which do not normalize their embeddings, the decomposition holds to within 0.006. Among the source documents, virtue ethics has the highest sentence-level coherence in every encoder (in SBERT, for example, $R$ = 0.602, against 0.572 for consequentialism and 0.561 for deontological ethics). Measures that retain the coherence factor, therefore, raise virtue ethics relative to consequentialism, and the reversal appears only where the gap between their average meanings is small: in SBERT, DistilBERT, E5-base, and GTE-base, the four encoders with the smallest consequentialism-virtue ethics gaps in Table 4. Deontological ethics never has the highest coherence of the three, yet it ranks first under every measure; its lead comes from the closeness of its average meaning to the Act, not from coherence or the choice of aggregation.

The comparisons so far treat each source document independently. Table 2, however, shows that the three normative theories share substantial content, so their associations with the target document cannot be assumed to be independent. To quantify this overlap, the source documents were also compared with one another under the same encoders.

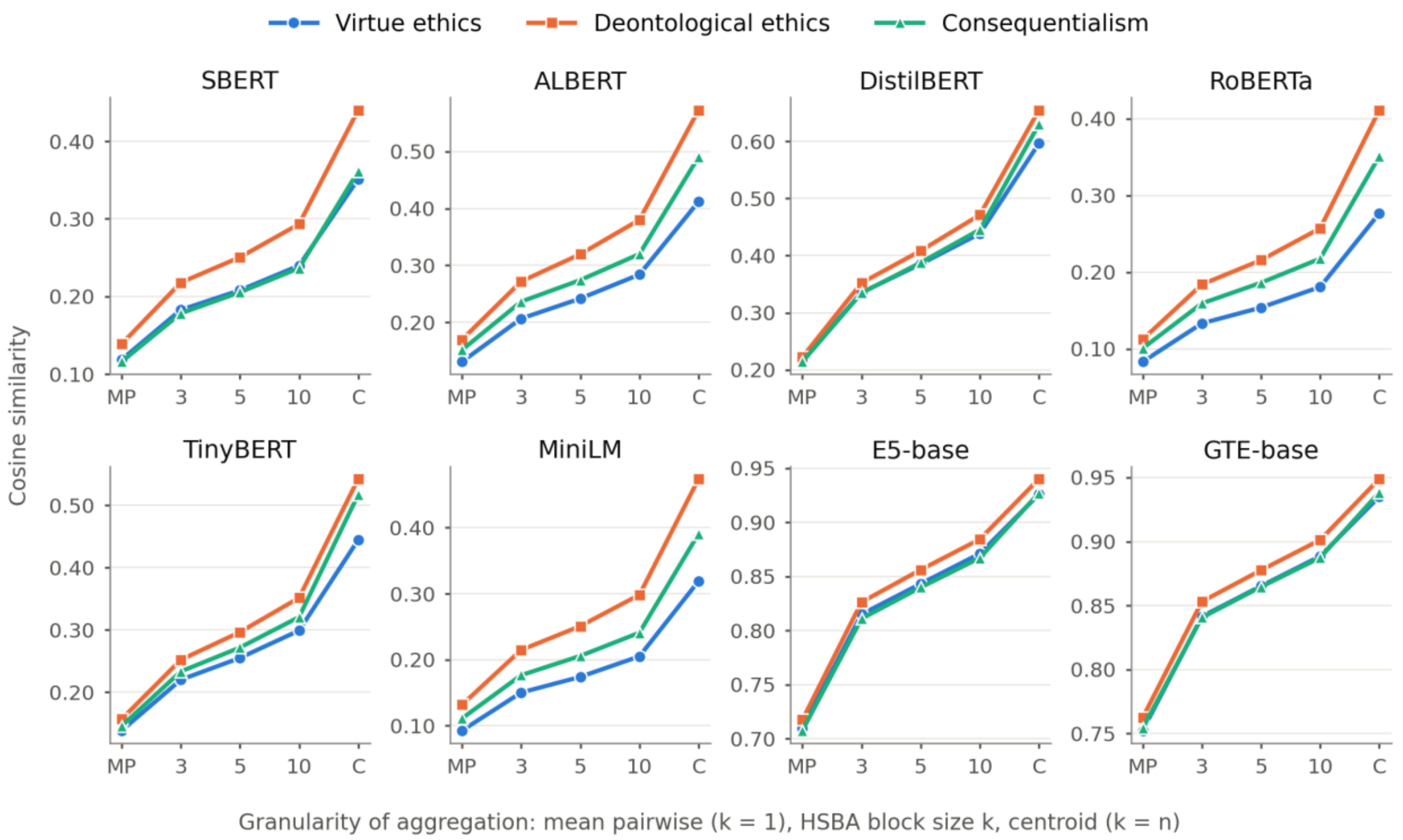

**(a)** Preamble of the EU AI Act.

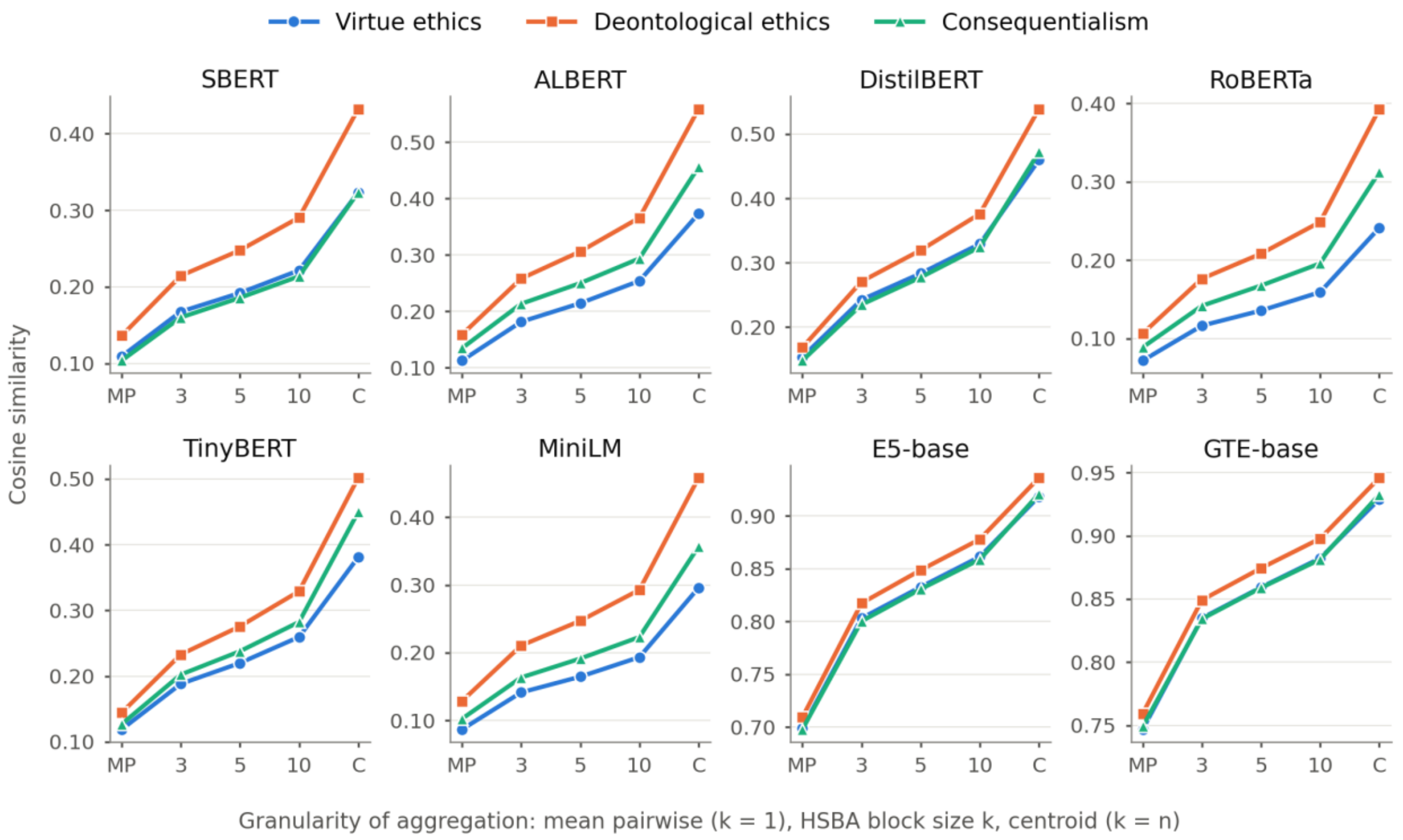

**(b)** Provisions of the EU AI Act.

**Figure 2** Source-target similarity by granularity of aggregation, from mean pairwise similarity (MP, $k = 1$) through HSBA ($k = 3, 5, 10$) to centroid similarity ($C, k = n$), for **(a)** the preamble and **(b)** the provisions of the EU AI Act. The vertical scale differs across panels because raw scores are comparable only within an encoder.

Table 6 reports the centroid and mean pairwise similarities between pairs of source documents. Under centroid similarity, deontological ethics and consequentialism form the most similar pair in all eight encoders. Mean pairwise similarity agrees in six encoders; DistilBERT and TinyBERT rank virtue ethics and consequentialism highest, again reflecting the higher coherence of the virtue ethics description. The ranking of the remaining two pairs varies across encoders.

Figure 3 compares these source-source similarities with the source-target similarities of Table 4. In every encoder, the least similar pair of source documents is more similar than the most similar source-target combination: for example, 0.765 against 0.440 in SBERT and 0.722 against 0.474 in MiniLM. Even in E5-base and GTE-base, with their compressed ranges, the source-source similarities (0.975 to 0.986) exceed all source-target similarities (0.918 to 0.949).

When explaining these findings, however, the encoders raise new questions rather than provide explanations, since the internal basis of each STS score is opaque. In contrast, the principles (algorithms), datasets, and identifiers employed by each encoder are known.

**Table 6** Similarity between pairs of source documents

| | Centroid cosine | | | Mean pairwise cosine | | |
|---|---|---|---|---|---|---|
| **Encoder** | **VE-DE** | **VE-CO** | **DE-CO** | **VE-DE** | **VE-CO** | **DE-CO** |
| *SBERT* | 0.789 | 0.765 | **0.861** | 0.267 | 0.263 | **0.276** |
| *ALBERT* | 0.776 | 0.808 | **0.897** | 0.243 | 0.261 | **0.271** |
| *DistilBERT* | 0.846 | 0.890 | **0.935** | 0.312 | **0.328** | 0.321 |
| *RoBERTa* | 0.772 | 0.765 | **0.873** | 0.233 | 0.242 | **0.251** |
| *TinyBERT* | 0.786 | 0.836 | **0.875** | 0.252 | **0.261** | 0.252 |
| *MiniLM* | 0.722 | 0.785 | **0.848** | 0.209 | 0.233 | **0.241** |
| *E5-base* | 0.975 | 0.975 | **0.983** | 0.747 | 0.748 | **0.750** |
| *GTE-base* | 0.977 | 0.976 | **0.986** | 0.790 | 0.790 | **0.797** |
| **Votes for the most similar pair** | 0 | 0 | 8 | 0 | 2 | 6 |

The highest score in each encoder and measure is in bold.

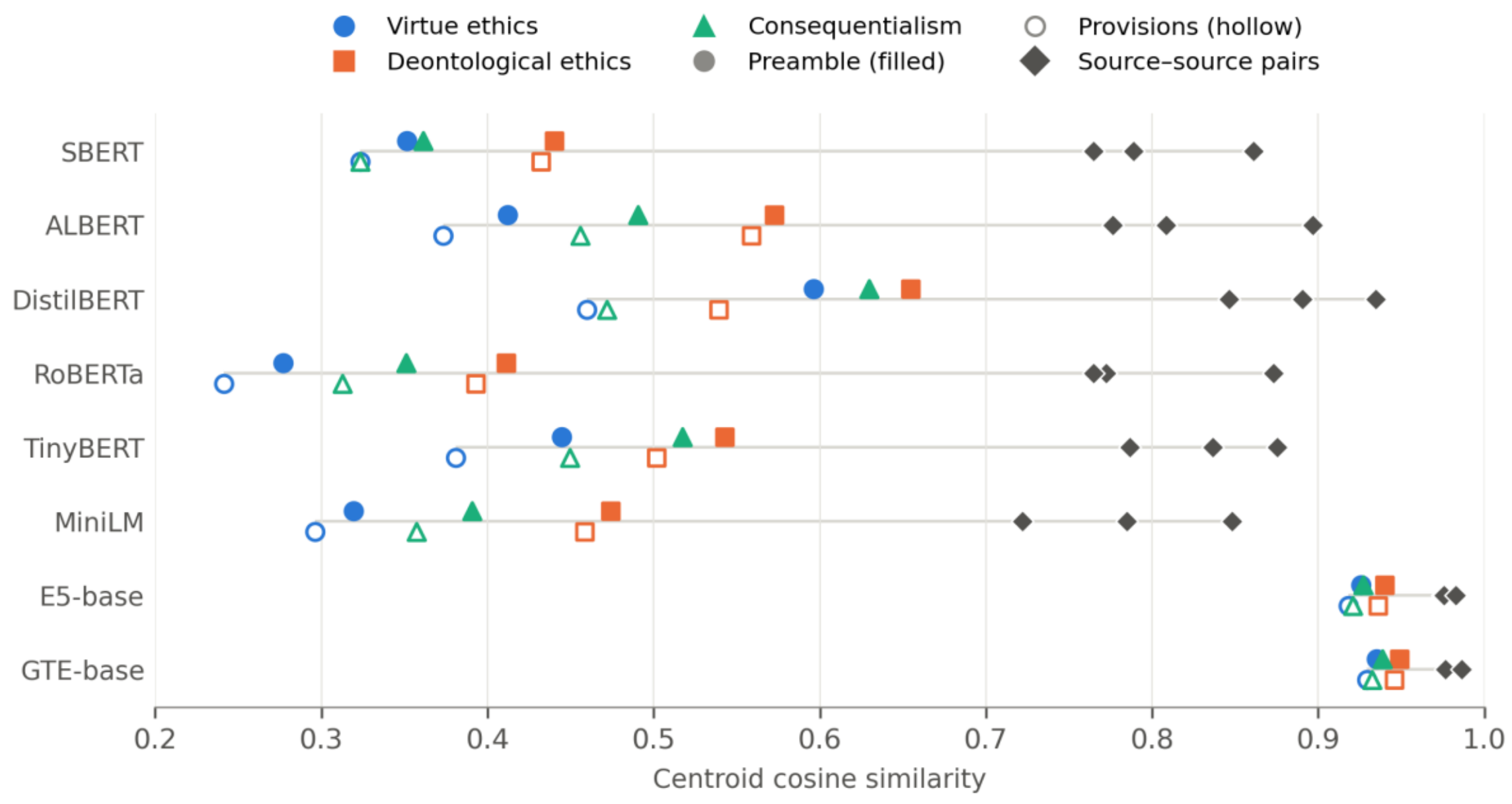

**Figure 3** Centroid cosine similarity of each source document with the preamble (filled markers) and the provisions (hollow markers), compared with the similarity between pairs of source documents (diamonds), by encoder

## 5. Discussion

The results reveal a consistent pattern: deontological ethics shows the strongest association with both target sections of the EU AI Act across all eight encoders and all three aggregation methods, from sentence pairs to whole documents. Consequentialism follows at the document level, and virtue ethics is third, although the order of these two depends on granularity. None of the source documents exhibits the degree of similarity they show with one another, which suggests that no single theory accounts for the ethical language of the Act. Instead, these results suggest that AI governance may be evolving into a novel form of "regulatory ethics" that selectively incorporates elements from multiple philosophical traditions to address the unique challenges posed by AI technologies.

The consistent lead of deontological ethics across both the preamble and provisions deserves deeper analysis. Deontological ethics is fundamentally rule-based, emphasizing duties, rights, and categorical principles. So are the legal texts. However, this raises an interpretive question: Does the higher similarity reflect genuine philosophical alignment with Kantian principles of human dignity and categorical imperatives or merely structural similarities between rule-based ethical systems and legal language? The EU AI Act's focus on stakeholder rights, banned practices, and compliance rules can create language patterns that align with duty-based deontological ethics, regardless of the underlying moral principles.

These results are consistent with the qualitative literature on the relationship between the normative theories and law. The three theories reach law at different levels: deontological ethics through the content of legislation, consequentialism through its design, and virtue ethics through its application by judges and officials (Kant, 1797/1996; Bentham, 1789/1996; Solum, 2003). A statutory text should therefore be most closely associated with deontological ethics, less so with consequentialism, and least with virtue ethics, which is the order observed here at the document level.

The low association of virtue ethics does not imply that the Act is indifferent to character or practical wisdom; the virtues on which it depends are exercised by those who apply it and are not stated in its text. Legal scholarship on the Act itself reads it as pursuing rights-based aims through a risk-based mechanism (Ho-Dac, 2025; Kaminski, 2023; Laux et al., 2024), and the difference between the target sections matches that reading: the preamble, where the Act states and justifies its purposes, is more closely associated with the theories of ethics than the provisions, where those purposes are translated into operational obligations. Because the two lines of evidence were obtained independently, one from the interpretation of the Act and its relationship to the theory of ethics and the other from the association of its language with that theory, their agreement strengthens both. The computational analysis also adds what the qualitative literature does not provide: a measure of how closely each theory is associated with each part of the Act, and of how robust that ordering is across encoders and levels of aggregation.

Future research should compare the EU AI Act with a substantial, systematically selected corpus of legislation to distinguish structural similarities inherent in legal language from substantive similarities specific to the Act, and to establish an appropriate legal-language baseline. Furthermore, the language of specific rights and duties should be freed from general rule-making structures, and the Act's stakeholder protections should be examined to determine whether they reflect genuine deontological principles or merely procedural compliance requirements.

The limited set of source texts representing theories of ethics—although its use is methodologically justified by Albayrakoglu and Aydin (2026)—remains a significant constraint. Each philosophical tradition spans centuries and encompasses multiple schools of thought that might yield different similarity scores.

The preprocessing of legal texts presents unique challenges that may have affected similarity calculations. Legal documents employ formatting conventions (article numbers, paragraph structures, cross-references) and syntactic patterns (dependent clauses, conditional statements, definitional sections). Treating complex legal sentences with multiple dependent clauses as separate units breaks down the logical structure of the legal reasoning. While necessary for computational processing, this approach may obscure the integrated argumentative patterns that characterize legal and philosophical discourse.

The encoders introduce further limitations. Their raw scores are not comparable across models, so the cross-encoder analysis relies on rank voting rather than averaging. Two of them, E5-base and GTE-base, return all source-target similarities within a narrow band near 0.93. A small share of long sentences, mostly in the preamble, exceeds the maximum input length of ALBERT and of the 128-token encoders and is truncated. However, the principal finding—the first-place ranking of deontological ethics—is unchanged among the encoders for which no sentence is truncated. Finally, the aggregation methods differ in what they measure: mean pairwise similarity and HSBA combine the association between documents with each document's internal coherence, whereas centroid similarity does not include this coherence factor. Rankings that hold under all three methods, such as the lead of deontological ethics, are therefore the most robust.

Table 6's revelation that the pairwise similarities between source documents exceed all source-target similarities creates another interpretive puzzle. This pattern suggests that the AI Act covers a wider ethical territory, encompassing traditional philosophical categories. It may also indicate a significant difference between legal language and philosophical discourse, leading to dissimilarities among related concepts. It may even be a symptom of poorly calibrated semantic similarity metrics used to analyze philosophical content.

Current pairwise methodologies cannot incorporate the source-source similarities shown in Table 6 into the source-target similarities in Table 4, because they treat each source document as an independent unit of comparison. Yet the three source documents overlap far more with one another than with the target document. A source-target score, therefore, mixes what a source document shares with the other sources and what it alone shares with the target, and pairwise STS offers no way to separate the two. Philosophical theories exist in discursive relationships with one another; ideas are repeatedly combined, critiqued, and modified for practical use, and legal texts draw on the resulting common stock rather than on any one theory in isolation. Measuring the association between a single source document and a target, conditional on the other source documents, requires a method that jointly models the sources.

Despite these methodological challenges, the findings offer valuable insights into the ethical foundations of AI governance. The consistent association between deontological ethics and the EU AI Act suggests that rights-based, duty-oriented approaches to AI ethics may have gained prominence in regulatory contexts. This finding has practical implications that depend on the stakeholders' power and attitudes. On the one hand, if AI regulation leans toward deontological frameworks, we might expect future governance approaches to emphasize categorical restrictions, inviolable rights, and duty-based compliance obligations. On the other hand, if it deviates from deontological frameworks, we might expect future governance approaches to emphasize consequentialist cost-benefit analyses, or, to a lesser extent, virtue-based professional ethics standards.

This discussion reveals that while STS analysis opens new possibilities for investigating the association between philosophical traditions and policy, it also generates interpretive challenges that require careful methodological development and theoretical sophistication to produce more specific results.

# 6. Conclusion

This study used NLP to gain insight into the association between ethics and law, drawing on a limited number of texts and eight Transformer-based sentence encoders. The results show that deontological ethics exhibits the highest semantic association with the EU AI Act among the theories examined. This finding is robust: it holds for all eight encoders, for both the preamble and the provisions, and at every level of aggregation from sentence pairs to whole documents.

The other two theories are associated with the Act to a lesser but comparable degree. At the document level, consequentialism ranks second and virtue ethics third, but this ordering is sensitive to the granularity of comparison and should be regarded as tentative. The preamble of the Act is more closely associated with all three theories than its provisions are.

This study should be regarded as a first attempt, and its caveats should be noted, as the number of texts on each theory needs to be increased to assess the degree of agreement on which theory is most closely associated with the Act. Therefore, several iterations of the work, using different yet comparable accounts of the theories, would yield more robust results.

Considering the Act, one potential issue is using its text as is. Whether keeping titles, paragraph numbers, or article numbers has distorted the results is unknown. The Act includes very long sentences, especially complex ones, in which each of the multiple dependent clauses appears as a distinct statement, each starting on a new line. This presents a challenge that may need to be addressed: repeating the independent clauses before each of their dependents does not appear to be a viable solution.

Methodologically, the study shows that the choice of aggregation method matters and can be made on principled grounds. Mean pairwise sentence similarity equals centroid similarity scaled by the coherence of the two documents, and hierarchical sentence-block aggregation spans the range between the two. Centroid similarity is therefore the appropriate primary measure of association between long documents, with the other two serving as checks on the robustness of rankings across granularities.

Aside from being compared to human judgment, the encoders in this study lack transparency, and none can justify its STS scores. Although they can yield STS scores between source documents, these scores cannot be justifiably incorporated into the source-target scores. There is a need for a new approach that accounts for overlap among source documents and determines each source document's association with the target, conditional on the others.

Future research in this area requires development along several dimensions: First, the corpus should be extended by incorporating larger, more representative collections of texts of ethics that capture the diversity within each theoretical tradition while maintaining comparability across theories. Second, transparent similarity metrics should be developed to identify the specific textual features that drive philosophical alignment, including vocabulary, argumentative structure, normative claims, and prescriptive language patterns. Third, joint modeling approaches represent perhaps the most important methodological frontier. There is an obvious need for feasible models that can account for the discursive relationships among philosophical theories and their joint association with legal texts, rather than treating each theory as an independent unit of comparison.

For the time being, this work's theoretical and methodological contribution to the literature surpasses its findings. Our work demonstrated NLP's ability to help uncover associations between texts that may have been lost over time or in translation. Even though our approach cannot establish how such associations arose, it can be used either to support and strengthen arguments about the relationship between bodies of text, as in the case of ethics and law handled in this work, or to develop hypotheses that may lead to further research, not only in ethics, but also in history, sociology, law, politics, literature, arts, etc.